\documentclass[12pt,a4paper]{article}%
\pdfoutput=1
\usepackage{amsmath,amssymb,amsfonts}
\usepackage[pdftex]{graphicx}
\usepackage{epsfig,url}
\usepackage{color}
\usepackage{float}
\usepackage[bf,footnotesize]{caption2}
\usepackage{graphicx}
\usepackage[]{hyperref}%
\numberwithin{equation}{section} 
\setcaptionmargin{1cm}

\def\dst{\displaystyle}
\def\f{\frac}
\def\({\left(}
\def\){\right)}

\begin{document}
\begin{titlepage}
\begin{flushright}
IFUP--TH/2010-26\hfill
\end{flushright}
\vskip 1.0cm
\begin{center}
{\Large \bf Signals of single particle production at the earliest LHC} \vskip 1.0cm
{\large Riccardo Barbieri$^{a,b}$ and Riccardo Torre$^{b,d}$}\\[0.7cm]
{\it $^a$ Scuola Normale Superiore, Piazza dei Cavalieri 7, I-56126 Pisa, Italy}\\[5mm]
{\it $^b$ INFN, Sezione di Pisa, Largo Fibonacci 3, I-56127 Pisa, Italy}\\[5mm]
{\it $^d$
Universit\`a di Pisa, Dipartimento di Fisica, Largo Fibonacci 3, I-56127 Pisa, Italy}
\end{center}
\vskip 1.0cm
\begin{abstract} Based on simple phenomenological Lagrangians, fulfilling reasonable consistency conditions, we consider under which circumstances the production of a single  particle  might be an early signal of new physics at the LHC. Effective final states are  $\gamma \gamma$ and  $\gamma +jet$ already with tens of inverse picobarns of integrated luminosity at 7 TeV.
\end{abstract}
\vskip 1cm \hspace{0.7cm} August 2010
\end{titlepage}

\section{Introduction}

The beginning of LHC operation at half the foreseen center-of-mass energy and with an unavoidably low initial luminosity brings the focus on possible signals of new physics that might show up at all in the early stages of operation. Model-building prejudices normally play an important role in determining the search strategies. While this is understandable, here we  set aside such prejudices as much as possible. We base our considerations on simple phenomenological Lagrangians, fulfilling reasonable consistency conditions. 
We are generally guided by the possible existence of composite states produced by a putative strong dynamics responsible for ElectroWeak Symmetry Breaking (EWSB), but we aim mostly at a neat and simple definition of the interactions responsible for the signals under discussion.
We think that this should also represent an appropriate guide  in the presentation of the experimental results, to maximize their usefulness for any subsequent consideration.

Single production of a relatively narrow resonance is the most obvious candidate for a copious source of new physics signals in the early stages of LHC operation. Very much studied is the case of a vector resonance produced in the $q \bar{q}$-channel, either neutral or charged, with its leptonic decay modes \cite{PDG,Salvioni:2009jp}. In fact, the multiplicity of  partonic channels in  a $p p$ collider makes several different cases possible. As a matter of fact, in the competition with the Tevatron with its integrated luminosity already available, the $q \bar{q}$-channel is definitely less favourable relative to the quark-quark, quark-gluon and gluon-gluon channels in view of the corresponding parton luminosities. We therefore concentrate our attention to these channels. If one considers in each of them the lowest possible spin and the lowest possible QCD representation, for matter of simplicity, the cases of interest are:
\begin{itemize}
\item {\large \bf $g g$-channel}: a spinless totally neutral scalar $S$;
\item {\large \bf $q g$-channel}: a $J=1/2$ colour-triplet "heavy quark", either "$U$" or "$D$";
\item {\large \bf $q q$-channel}: a spinless colour-triplet or colour-sextet $\phi$, with various possible charges.
\end{itemize}
We assume that one single new particle is available at a time, which therefore can only decay into Standard Model (SM) particles. We concentrate our attention to the first two cases, since the scalar triplet, or sextet\cite{Bauer:2009cc}, can only decay into a pair of jets, consistently with known constraints, whereas we find relatively more promising the final states containing at least one photon. The resonance in the $q q$ channel suffers also of problems with flavour physics (see below).

\section{Neutral scalar singlet $S$}

The reference Lagrangian that we adopt is the following,
\begin{equation}
\mathcal{L}_{S}=c_{3}\frac{g_{S}^{2}}{\Lambda}G_{\mu\nu}^{a}G^{\mu\nu\,a}S+c_{2}\frac{g^{2}}{\Lambda}W_{\mu\nu}^{i}W^{\mu\nu\,i}S+c_{1}\frac{g^{\prime 2}}{\Lambda}B_{\mu\nu}B^{\mu\nu}S+\sum_{f}c_{f}\frac{m_{f}}{\Lambda}\bar{f}fS\,,
\label{LS}
\end{equation}
where $\Lambda$ is an energy scale, the $c_i,~i=1,2,3,f$, are dimensionless coefficients and $f$ is any SM fermion, of mass $m_f$. $S$ is a scalar of mass $M_S$, totally neutral under the SM gauge group. A coupling of $S$ to the Higgs doublet of the form $m_S S H^+ H$ could also be present, as it would actually be induced by radiative corrections. In a wide range of $M_S$ and $\Lambda$ it is however consistent to assume that such coupling is irrelevant.

\begin{figure}[htbp]
\begin{center}
\includegraphics[scale=0.35]{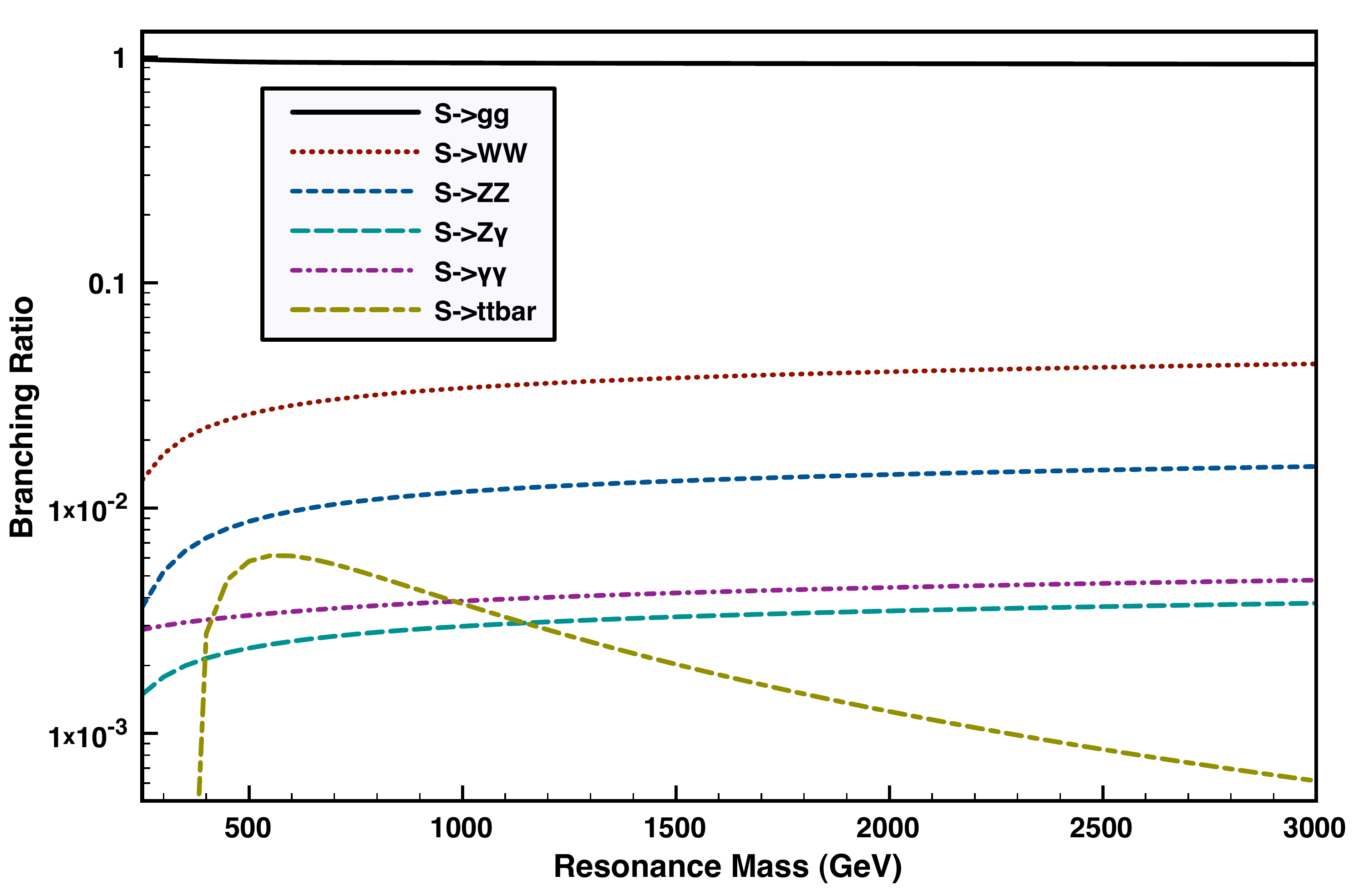}
\vspace{-2mm}\caption{Branching ratios as functions of the scalar mass $M_{S}$ for $c_i=1$.}
\label{BR_scalar}
\end{center}
\end{figure}

The two body widths of $S$ are given by
\begin{subequations}\label{eq2}
\begin{align}
&\dst \Gamma\(S\to gg\)=\f{2c_{3}^{2}g_{S}^{4}M_{S}^{3}}{\pi\Lambda^{2}}\,,\\
&\dst \Gamma\(S\to W^{+}W^{-}\)=\f{c_{2}^{2}g^{4}\sqrt{M_{S}^{2}-4M_{W}^{2}}\(M_{S}^{4}-4M_{S}^{2}M_{W}^{2}+6M_{W}^{4}\)}{2\pi M_{S}^{2}\Lambda^{2}}\,,\\
&\dst \Gamma\(S\to ZZ\)=\f{\(c_{2}g^{2}\cos^{2}\theta_{W}+c_{1}g^{\prime 2}\sin^{2}\theta_{W}\)^{2}\sqrt{M_{S}^{2}-4M_{Z}^{2}}\(M_{S}^{4}-4M_{S}^{2}M_{Z}^{2}+6M_{Z}^{4}\)}{4\pi M_{S}^{2}\Lambda^{2}}\,,\\
&\dst \Gamma\(S\to Z\gamma\)
=\f{\sin^{2}\theta_{W}\cos^{2}\theta_{W}\(c_{1}g^{\prime 2}-c_{2}g^{2}\)^{2}\(M_{S}^{2}-M_{Z}^{3}\)^{2}}{2\pi M_{S}^{3}\Lambda^{2}}\,,\\
&\dst \Gamma\(S\to \gamma\gamma\)=\f{\(c_{1}+c_{2}\)^{2}e^{4}M_{S}^{3}}{4\pi\Lambda^{2}}\,,\\
&\dst \Gamma\(S\to t\bar{t}\)=\f{3c_{t}^{2}m_{t}^2\(M_{S}^{2}-4m_{t}^{2}\)\sqrt{M_{S}^{2}-4m_{t}^{2}}}{8\pi M_{S}^{2}\Lambda^{2}}\,,
\end{align}
\end{subequations}
with the corresponding Branching Ratios shown in Fig. \ref{BR_scalar} as functions of $M_S$ for all $c_i=1$, which we take as reference case. 
The total width of $S$ scales as $1/\Lambda^2$. $\Lambda$ is an arbitrary mass parameter, which we take for illustrative purposes at 3 TeV, reminiscent of a  new strong interaction at $\Lambda \approx 4\pi v$, possibly responsible for EWSB. If $S$ were a composite particle in (\ref{LS}) by the strong dynamics, Naive Dimensional Analysis \cite{Manohar:1983md,Cohen:1997rt} (NDA) would suggest $c_i\approx 1/4\pi$. Larger values could however arise from large N and/or from more drastic assumptions about the nature of the gauge bosons.
 For $\Lambda = 3$ TeV and $c_i = 1$, $\Gamma_S$ goes from about 10 GeV at $M_S= 500$ GeV to about 250 GeV for $M_S=1.5$ TeV.

\begin{figure}[ptbh]
\begin{minipage}[b]{8.2cm}
\centering
\includegraphics[scale=0.25]{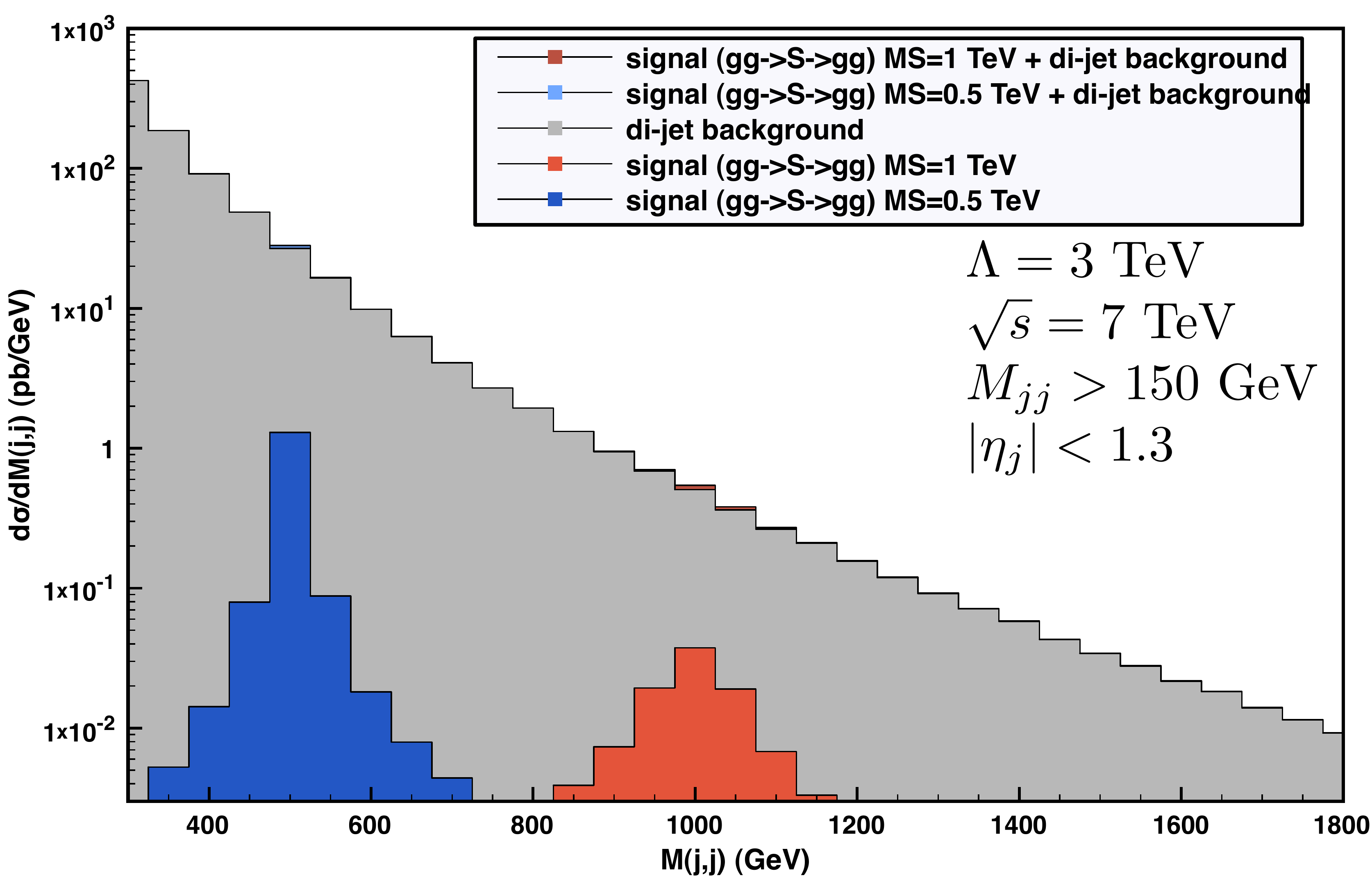}
\end{minipage}
\ \hspace{2mm} \hspace{3mm} \ \begin{minipage}[b]{8.5cm}
\centering
\includegraphics[scale=0.25]{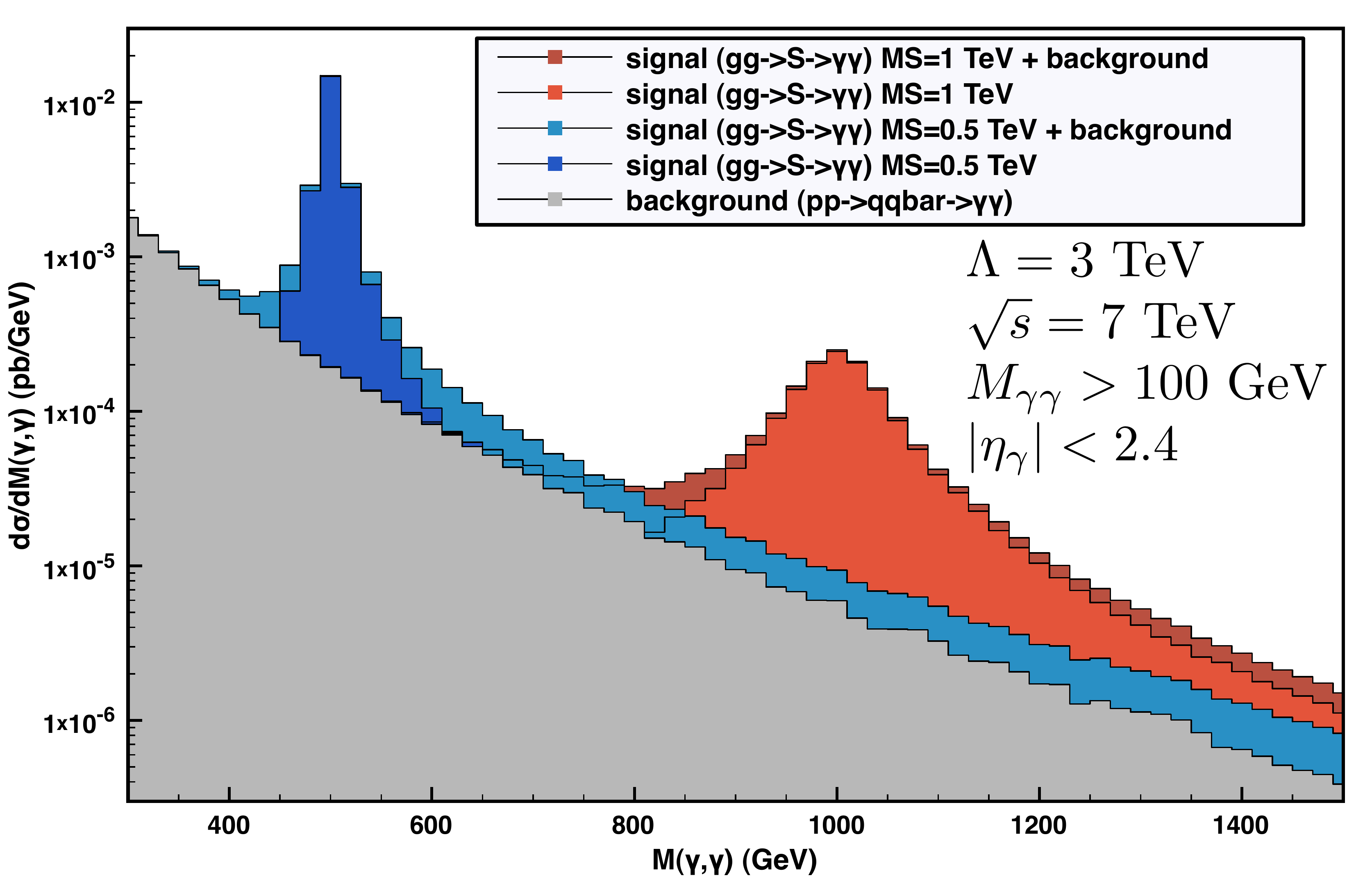}
\end{minipage}
\caption{Signals of a $500$ GeV (blue) and $1$ TeV (red) scalar singlet $S$ in the di-$jet$ (left panel) and $\gamma\gamma$ (right panel) invariant mass distributions vs the SM background at the early LHC ($\sqrt{s}=7$ TeV). $ \Lambda = 3$ TeV, $c_i=1$.}%
\label{Scalar_1}%
\end{figure}

\begin{figure}[ptbh]
\begin{minipage}[b]{8.2cm}
\centering
\includegraphics[scale=0.25]{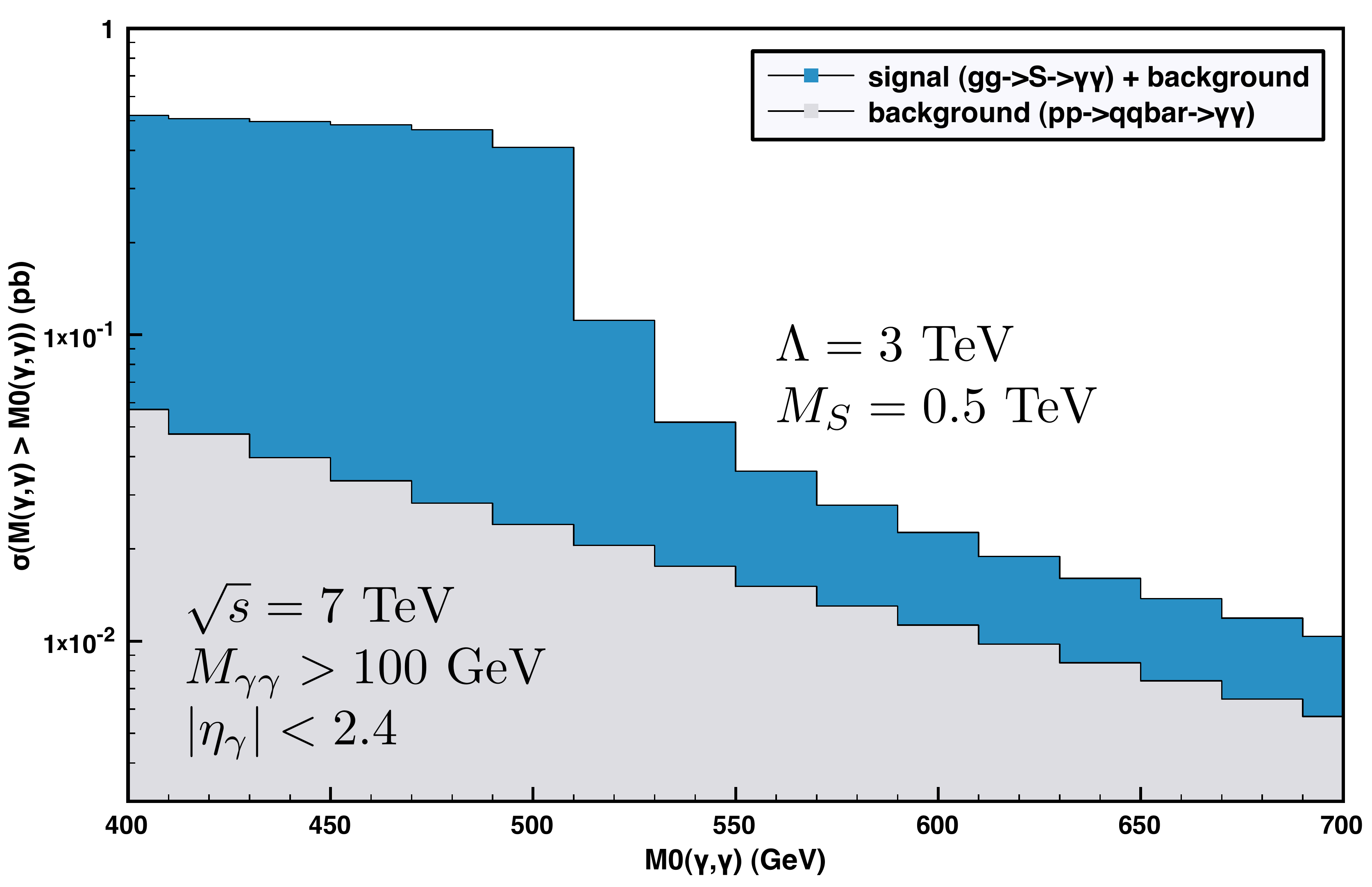}
\end{minipage}
\ \hspace{2mm} \hspace{3mm} \ \begin{minipage}[b]{8.5cm}
\centering
\includegraphics[scale=0.25]{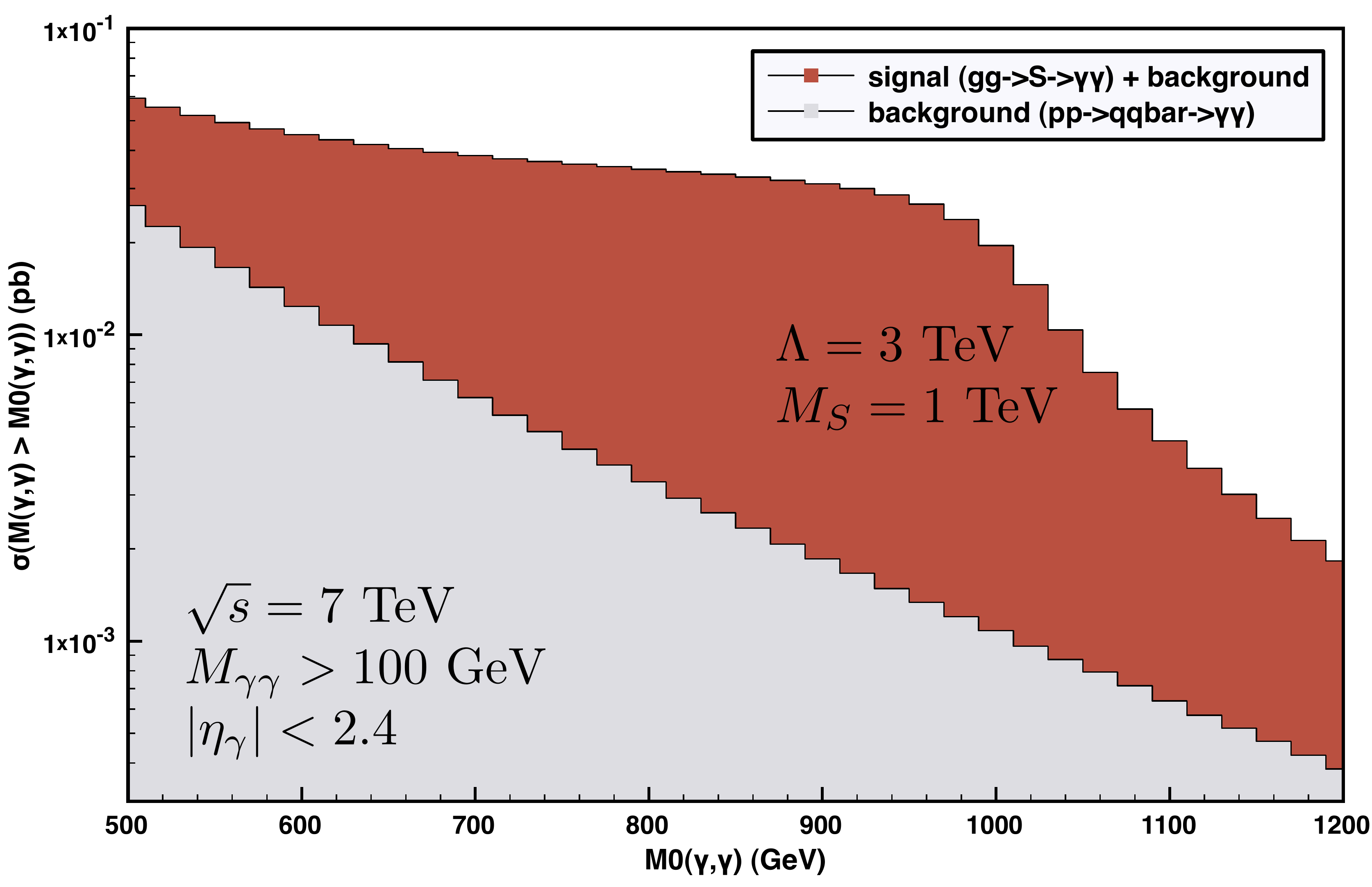}
\end{minipage}
\caption{ Number of events for $\int{\mathcal{L} dt} = 1$ pb$^{-1}$ at LHC (7 TeV) with $\gamma \gamma$ invariant mass greater than $M_{0}(\gamma, \gamma)$ for a $500$ GeV (left panel) and a $1$ TeV (right panel) scalar singlet $S$. $\Lambda = 3$ TeV, $c_i=1$.}
\label{Scalar_2}%
\end{figure}

We have made a preliminary study of the sensitivity to the search 
for $S$ in the di-$jet$, $\gamma\gamma$ and $\gamma Z$ channels. The results for the first two cases are illustrated in Fig. \ref{Scalar_1} for $M_S=0.5$ and $M_S=1$ TeV respectively. While it appears difficult to see an emerging signal in the di-$jet$ final state, taking into account of systematic uncertainties, a discovery  looks possible in the $\gamma\gamma$ channel with a modest luminosity of about $10$ to $100$ pb$^{-1}$, as illustrated in Fig. \ref{Scalar_2}. In the case of $\gamma Z$ channel one would have to pay for an extra factor of about $20\div40$ in the needed luminosity.

All this is trivially extendible to arbitrary values of $\Lambda$ and $c_i$. For a given integrated luminosity the absence of signal in  $\gamma \gamma$, assuming the dominance of $S\rightarrow gg$, is easily translated in a lower bound on $\Lambda/(c_1 + c_2)$.

\section{Heavy quark $U$}

In analogy with \ref{LS}, to describe the interactions of a heavy $U$-type quark  of mass $M_U$ with the SM particles we consider the interaction Lagrangian
\begin{equation}\label{LU}
\mathcal{L}_{U}=c_{G}\frac{g_{S}}{\Lambda}\bar{U}_{L}\sigma^{\mu\nu} T^{a} u_{R} G_{\mu\nu}^{a}+c_{B}\frac{g^{\prime}}{\Lambda}\bar{U}_{L}\sigma^{\mu\nu} u_{R} B_{\mu\nu}+\text{h.c.}\,,
\end{equation}
where $\sigma^{\mu\nu}=i/2 [\gamma^{\mu},\gamma^{\nu}]$ and $T^{a}$ are the generators of the  fundamental representation of SU(3) ($T^{a} = \lambda^a/2$). 
$U$ transforms therefore as a $(3,1)_{2/3}$ of the SM gauge group. 
As in the case of the scalar $S$, $U$ could be a composite state by a strong dynamics responsible for EWSB, in which case NDA suggests $\Lambda \approx 4\pi v \approx 3$ TeV and $c_G \approx c_B \approx 1/4\pi$.

Note that the Lagrangian \ref{LU} does not break the chirality of the standard $u$-quark, which is crucial to preserve its lightness. A problem with flavour arises, however, since we pretend that $U$ couples to the physical standard $u$-quark but has no (significant) coupling to the $c$-quark, not to cause unobserved $\Delta C=2$ flavour changing effects. In other words the coupling in (\ref{LU}) would have to be particularly aligned with the standard Yukawa matrix for the up-type quarks. While this is not excluded, a neater way to get around this problem would be to introduce three $U$-fields, one per generation, and assume that the Lagrangian (\ref{LU}) respects a global $SU(3)_R$ acting on the standard three up-type quarks. In either case the following considerations apply\footnote{Analogous considerations hold for the scalar states in the $q q$ channel. For example in the case of the $u u$ channel one would have to introduce a $SU(3)_R$-sextet of highly degenerate $\phi$'s.}.

\begin{figure}[ptbh]
\begin{minipage}[b]{8.2cm}
\centering
\includegraphics[scale=0.25]{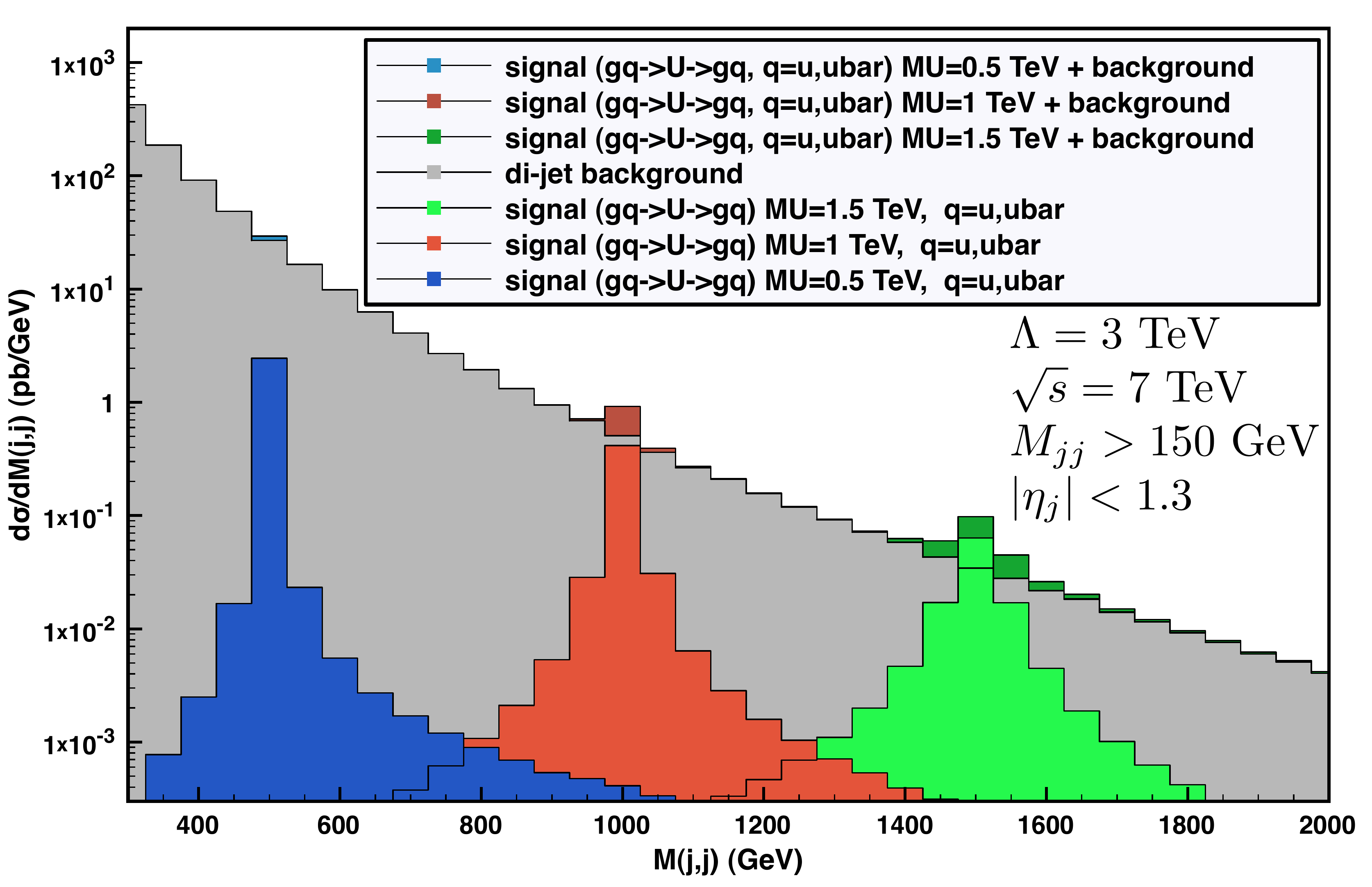}
\end{minipage}
\ \hspace{2mm} \hspace{3mm} \ \begin{minipage}[b]{8.5cm}
\centering
\includegraphics[scale=0.25]{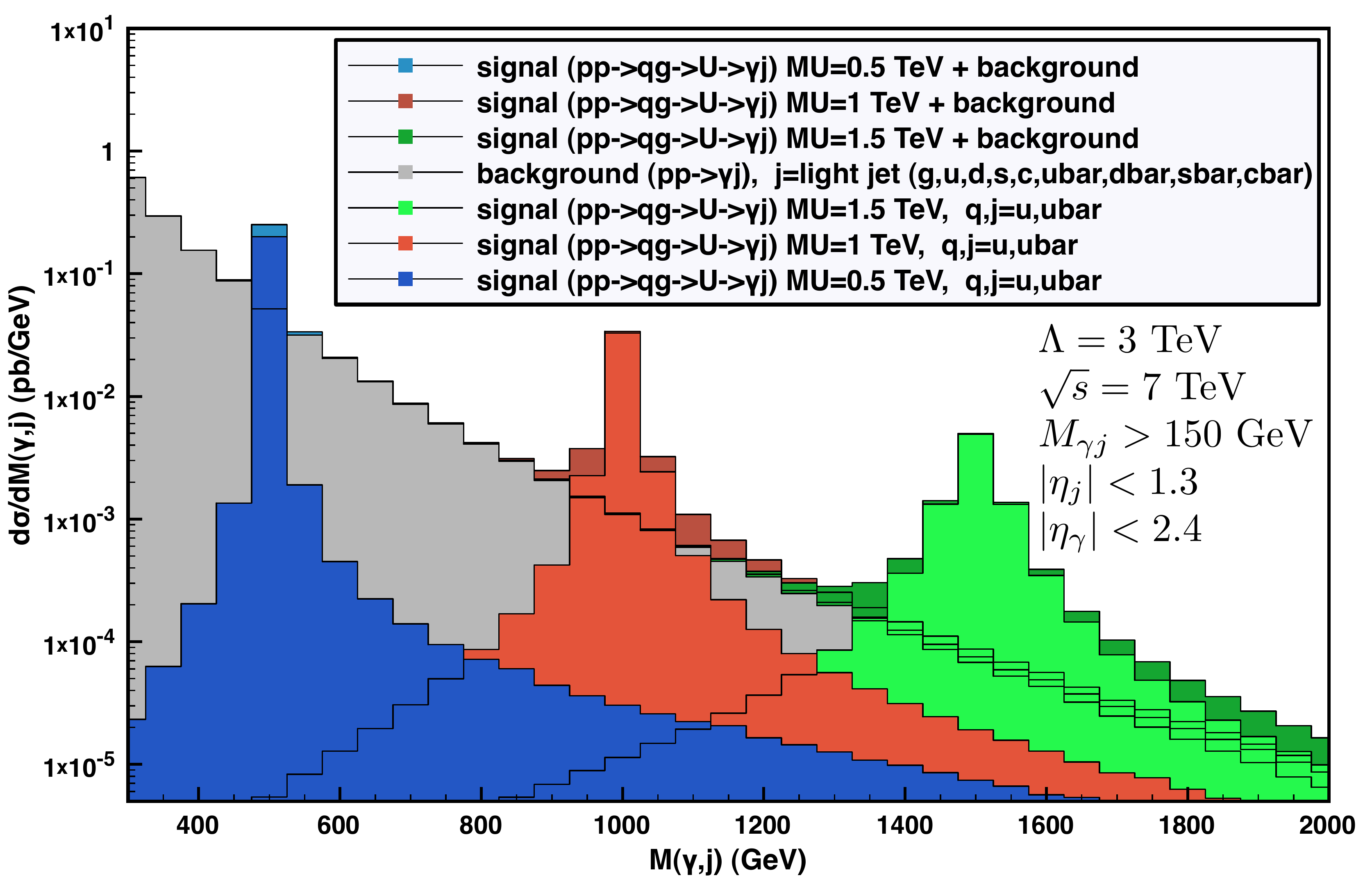}
\end{minipage}
\caption{Signals of a $500$ GeV (blue), $1$ TeV (red) and $1.5$ TeV (green) heavy quark in the di-$jet$ (left panel) and $\gamma+jet$ (right panel) invariant mass distribution vs the SM background at the early LHC ($\sqrt{s}=7$ TeV). $\Lambda = 3$ TeV, $c_i=1$.}%
\label{Fermion_1}%
\end{figure}

To study the sensitivity to $U$, as in the case of the singlet $S$ we consider the reference values $\Lambda=3$ TeV and $c_G = c_B =1$. The two body widths of $U$ are given by
\begin{subequations}\label{eq5}
\begin{align}
&\dst \Gamma\(U\to ug\)
=\f{4\alpha_{S}M_{U}^{3}}{3\Lambda^{2}}\,,\\
&\dst \Gamma\(U\to uZ\)
=\f{g^{\prime\,2}\sin^{2}\theta_{W}\sqrt{M_{U}^{2}-4M_{Z}^{2}}\(2M_{U}^{4}-M_{Z}^{2}M_{U}^{2}-M_{Z}^{4}\)}{8\pi M_{U}^{2}\Lambda^{2}}\approx \f{\alpha M_{U}^{3}\tan^{2}\theta_{W}}{\Lambda^{2}}\,,\\
&\dst \Gamma\(U\to u\gamma\)
=\f{\alpha M_{U}^{3}}{\Lambda^{2}}\,,
\end{align}
\end{subequations}
so that the total width,
\begin{equation}
\Gamma_U \approx \frac{M_{U}^{3}\alpha_{S}}{\Lambda^{2}} (\frac{4}{3}
+\frac{\alpha}{\alpha_{S}\cos^{2}\theta_{W}})\,,
\end{equation}
ranges from about 2 GeV to about 30 GeV for $M_U=0.5$ to $M_U=1.5$ TeV. Almost irrespective of the $U$-mass, 
the dominant $u g$ decay mode has a Branching Ratio of about 92$\%$, whereas the $u \gamma$ and $u Z$ modes have BRs of about 6$\%$ and 2$\%$ respectively.

\begin{figure}[ptbh]
\begin{minipage}[b]{8.2cm}
\centering
\includegraphics[scale=0.25]{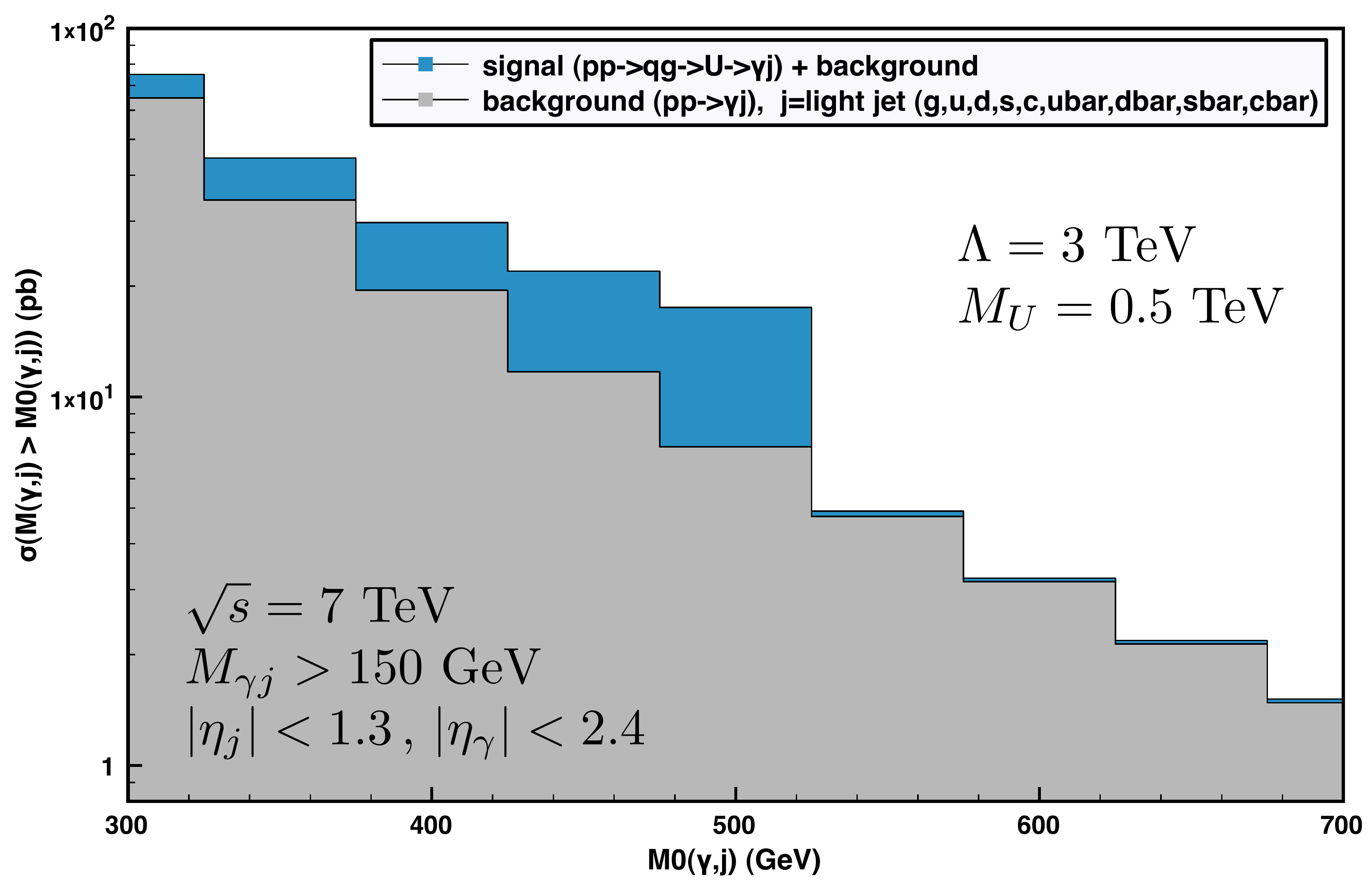}
\end{minipage}
\ \hspace{2mm} \hspace{3mm} \ \begin{minipage}[b]{8.5cm}
\centering
\includegraphics[scale=0.25]{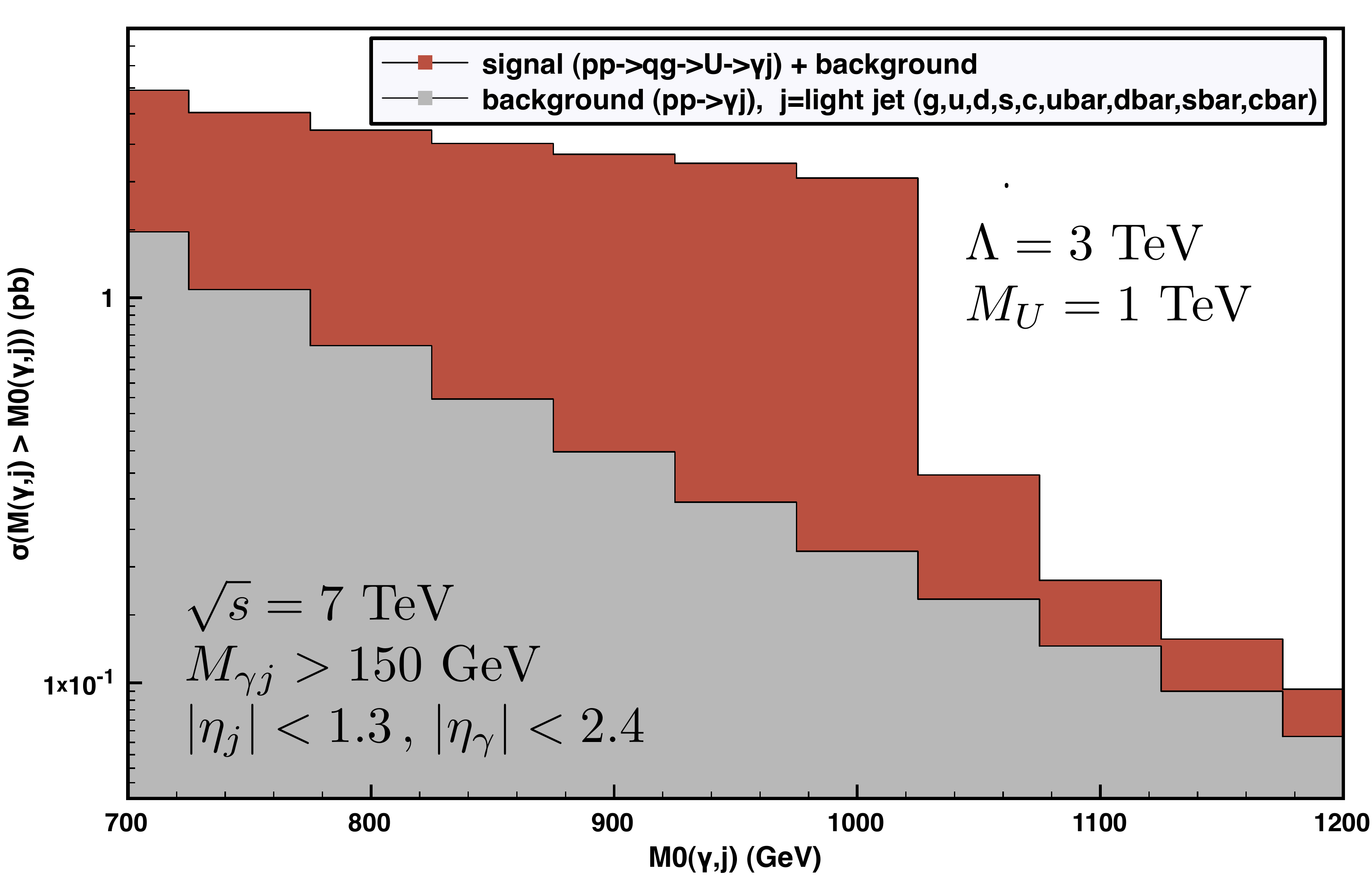}
\end{minipage}
\ \hspace{2mm} \hspace{3mm} \ \begin{minipage}[b]{8.5cm}
\centering
\includegraphics[scale=0.25]{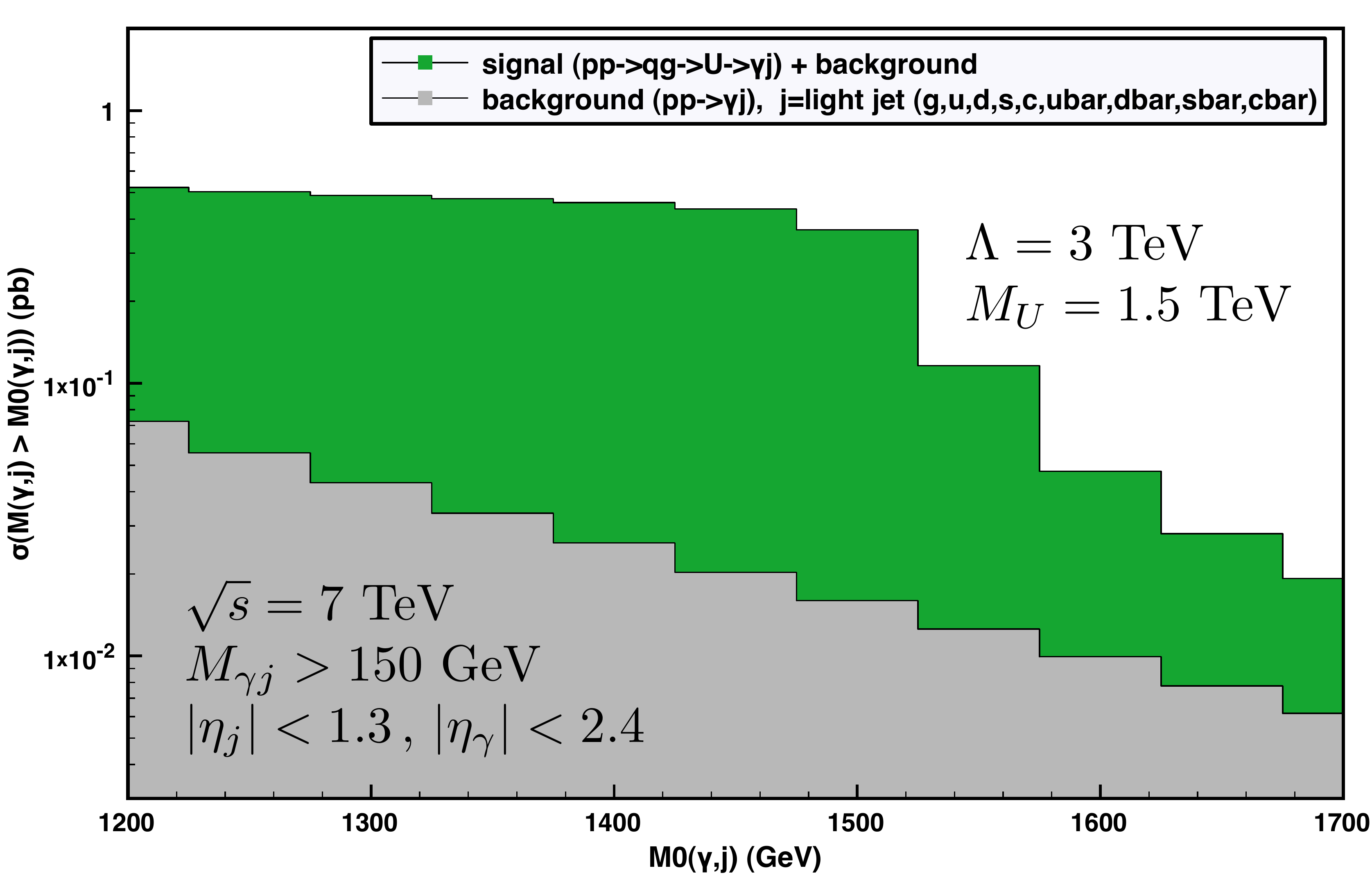}
\end{minipage}
\caption{Number of events for $\int{\mathcal{L} dt} = 1$ pb$^{-1}$ at LHC (7 TeV) with $\gamma + jet$ invariant mass greater than $M_{0}(\gamma, j)$ for a $500$ GeV (top left panel), a $1$ TeV (top right panel) and a $1.5$ TeV (bottom left panel) heavy quark. $\Lambda = 3$ TeV, $c_i=1$.}
\label{Fermion_2}%
\end{figure}

As in the case of $S$, we have made a preliminary study of the sensitivity to the search 
for the heavy $U$ in the di-$jet$, $\gamma+jet$ and $Z+jet$ channels. The results for the first two cases are illustrated in Fig. \ref{Fermion_1} for $M_U=0.5,~1$ and 1.5 TeV respectively. 
While it appears difficult to see an emerging signal in the di-$jet$ final state, taking into account of systematic uncertainties, a discovery  looks possible in the $\gamma +jet$ channel with a modest luminosity of about $5$ or $10$ pb$^{-1}$, as illustrated in Fig. \ref{Fermion_2}.
The $Z+jet$ channel  would require luminosities in the hundreds of inverse picobarns.

\section{Conclusions}

The lack so far of a thorough experimental exploration of the energy range at or well above the Fermi scale, the most fundamental scale in particle physics as presently known,  calls for 
an open attitude in the expectation for possible signals of new physics. In turn this has a twofold general implication. On one side one must be ready for surprises. On the other side it is essential that the experimental results be analyzed 
and presented in the neatest possible way with a minimum of biases.

In this work we have attempted to apply this attitude to the discussion of possible signals of new physics in the early stages of LHC operation, concentrating our attention to the production of a relatively narrow resonance that can compete with the well studied case of a neutral vector. While we have generally in mind a possible composite state produced by a putative strong dynamics responsible for EWSB, the important thing is that the interaction that produces one of these states be clearly and simply defined. This is the case for the phenomenological Lagrangians 
(\ref{LS}) and (\ref{LU}) with a single interaction term responsible for the production for the singlet scalar $S$ or for the triplet fermion $U$ respectively. 

With a suitable guess for the decay branching ratios we have found that final states in a pair of photons for $S$ or in a photon plus jet for $U$ may lead to detectable signals already with a few tens of picobarns of integrated luminosity at 7 TeV. For this to be the case we may have used an optimistic value for the strength of the relevant couplings. While such values cannot in any case be excluded a priori and, to the best of our knowledge, have not been excluded at the Tevatron so far, the progression of the integrated luminosity achievable at the LHC can quickly explore fully relevant regions for  these couplings. Meanwhile, with a defined and simple single coupling for the production of $S$ or $U$,  the experimental results for a narrow resonance in the $\gamma \gamma$ or in the $\gamma+ jet$ channels can be usefully given as a plot of $d\sigma/dM\cdot BR$ versus the invariant mass of the final state channel. We think that the interpretation of a possible positive signal in terms of more elaborate and more defined theoretical models would be eased by this way of presenting the data.

\subsection*{Acknowledgements}

{We thank Michelangelo Mangano, Riccardo Rattazzi and Slava Rychkov for useful comments.
This work is supported in part by the European Programme ``Unification in the LHC Era",  contract PITN-GA-2009-237920 (UNILHC) and by the
MIUR under the PRIN contract 2008XM9HLM.}

\end{document}